\documentclass[11pt,twoside]{article}

\usepackage{asp2006}
\usepackage{lscape}
\usepackage{graphicx}
\usepackage{amssymb}

\begin{document}
\title{Two Suggestions to See the Hidden Magnetism of the Solar Chromosphere}   
\author{J. Trujillo Bueno}   
\affil{Instituto de Astrof\'{\i}sica de Canarias, 38205, La Laguna, Tenerife, Spain}    

\begin{abstract} 
Solar magnetic fields leave their fingerprints in the polarization signatures of the emergent spectral line radiation. This occurs through a variety of rather unfamiliar physical mechanisms, not only via the Zeeman effect. In particular, magnetic fields modify the atomic level polarization (population imbalances and quantum coherences) that anisotropic radiative pumping processes induce in the atoms and molecules of the solar atmosphere. Interestingly, this so-called Hanle effect allows us to ``see" magnetic fields to which the Zeeman effect is blind within the limitations of the available instrumentation. Here I argue that the Ca~{\sc ii} IR triplet and the He~{\sc i} 10830~\AA\ multiplet would be very suitable choices for investigating the magnetism of the solar chromosphere via spectropolarimetric observations from a future space telescope, such as JAXA's SOLAR-C mission.
\end{abstract}


\section{Introduction}

We may define ``the Sun's hidden magnetism" as all 
the magnetic fields of the extended solar atmosphere that are impossible 
to diagnose via the only consideration of the Zeeman effect. 
Contrary to what one might think, 
there are many examples that belong to this category:

\begin{itemize}

\item Most of the magnetism of the quiet solar photosphere.

\item The magnetic fields of the solar chromosphere outside sunspots, 
including the spike-like jet features that we call spicules.

\item The magnetic fields that confine the plasma of solar prominences and 
filaments, including those of active regions.

\item The magnetism of the solar transition region and corona.

\end{itemize}

The reasons are the following, but see more information in the paper by Trujillo Bueno (2009) on {\em Recent Advances in Chromospheric and Coronal Polarization Diagnostics}. First, the polarization of the Zeeman effect as a diagnostic
tool is blind to magnetic fields that are randomly oriented on scales too small to be resolved. Second, the circular polarization induced by the longitudinal Zeeman effect scales with the ratio, ${\cal R}$, between the Zeeman splitting and the Doppler broadened line width, while the Stokes $Q$ and $U$ signals produced by the transverse Zeeman effect scale as ${\cal R}^2$. Therefore, the Zeeman effect is of limited practical interest for the exploration of magnetic fields in hot (chromospheric and coronal) plasmas.

Fortunately, there is another physical mechanism by means of which 
the magnetic fields of the solar atmosphere leave fingerprints 
in the polarization of the emergent spectral line radiation: the Hanle effect.
Anisotropic radiation pumping processes produce atomic level polarization (i.e., population imbalances and quantum coherences among the magnetic sublevels pertaining to any given degenerate energy level). The Hanle effect can be defined as any modification of the atomic level polarization due to the presence of a magnetic field, including the remarkable effects produced by the level crossings and repulsions that take place when going from the Zeeman effect regime to the complete Paschen-Back effect regime (e.g., Landi Degl'Innocenti \& Landolfi 2004). 
The Hanle effect is especially sensitive to magnetic strengths between $0.1\,B_H$ and $10\,B_H$, where the critical Hanle field intensity ($B_H$, in gauss) is that for which the Zeeman splitting of the $J$-level under consideration is similar to its natural width: $B_{\rm H}=(1.137\times10^{-7})/(t_{\rm life}\,g_J)$ (with $g_J$ the level's Land\'e factor and $t_{\rm life}$ its radiative lifetime in seconds). Since the lifetimes of the upper levels ($J_u$) of the transitions of interest are usually much smaller than those of the lower levels ($J_l$), clearly diagnostic
techniques based on the lower-level Hanle effect are sensitive to much
weaker fields than those based on the upper-level Hanle effect. Summarizing:

\begin{itemize}

\item The Hanle effect is sensitive to weaker magnetic fields than the Zeeman effect: from at least 1~mG to a few hundred gauss. Moreover, it is sensitive to magnetic fields that are randomly oriented on scales too small to
be resolved (e.g., the Hanle-effect investigation by Trujillo Bueno et al. (2004) showed that the bulk of the ``quiet" photosphere is teeming with tangled magnetic fields at subresolution scales, with 
$\langle B \rangle{\sim}100$ G, which 
support the suggestion that a solar surface dynamo plays 
a significant role for the quiet Sun magnetism). 

\item The Hanle effect as a diagnostic tool is {\em not\/} limited to a narrow solar limb
zone. In particular, in the forward scattering geometry of a solar disk center observation,
the Hanle effect creates linear polarization in the presence of an  
inclined magnetic field (e.g., Trujillo Bueno et al. 2002).

\end{itemize}
 
The disadvantage of the Hanle effect as a diagnostic tool 
is that for magnetic strengths 
$B > 10\,B_{\rm H}(J_u)$ the linear polarization signals are
sensitive only to the orientation of the magnetic field vector. Fortunately, the Hanle and Zeeman effects can be suitably complemented for exploring magnetic fields in solar and stellar physics. 

\section{How to explore the magnetic fields of the solar chromosphere?}

There are several possibilities for mapping the magnetic fields of the solar chromosphere, 
but the following spectral lines would be very suitable choices to guarantee important scientific discoveries via spectropolarimetric observations from a future space telescope, like JAXA's SOLAR-C mission.

\subsection{The IR triplet of Ca~{\sc ii}}

We already know that diagnostic techniques based on the Zeeman effect in the
Ca~{\sc ii} IR triplet are very useful for obtaining information on the three-dimensional 
structure of sunspots magnetic fields 
(e.g., Socas-Navarro et al. 2000; Socas-Navarro 2005). 
But, what Stokes profiles do we see outside sunspots?

Figure 1 shows a high-sensitivity
spectropolarimetric observation of the quiet solar chromosphere in
the strongest (8542~\AA) line of the Ca~{\sc ii} triplet. It was obtained
by R. Ramelli (IRSOL), R. Manso Sainz (IAC) and me using the
Z\"urich Imaging Polarimeter (ZIMPOL) attached to THEMIS. The
observed Stokes $V/I$ profiles are clearly caused by the longitudinal
Zeeman effect, but the Stokes $Q/I$ and $U/I$ signals are produced
mainly by the influence of atomic level
polarization. Although
the spatio-temporal resolution of this spectropolarimetric observation
is rather low (i.e., no better than $3\arcsec$ and 20 minutes), the
fractional polarization amplitudes fluctuate between 0.01\% and 0.1\%
along the spatial direction of the spectrograph's slit, with a typical
spatial scale of $5\arcsec$. Note that while the Stokes
$Q/I$ signal changes its amplitude but remains always positive along
that spatial direction, the sign of the Stokes $U/I$ signal
fluctuates.

\begin{figure}  
  \centering
  \includegraphics[width=11.5cm]{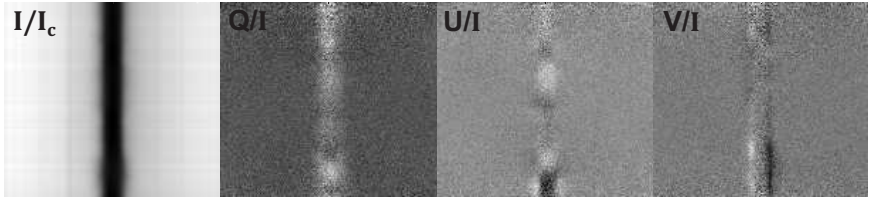}
\caption[]{
  An example of our recent spectropolarimetric observations of the
  Ca~{\sc ii} 8542~\AA\ line in a very quiet region close to the solar
  limb, using ZIMPOL at the French-Italian telescope THEMIS. 
  The reference direction for Stokes
  $Q$ is the tangent to the closest limb. From Trujillo Bueno et al. (2009).
}\end{figure}

The physical interpretation of these spectropolarimetric
observations requires solving the so-called 
NLTE problem of the $2^{\rm nd}$ kind (e.g., via the application of the multilevel radiative transfer code  
MULTIPOL described in Manso Sainz \& Trujillo Bueno 2003a). Interestingly, 
while the scattering polarization in the 8498 \AA\ line shows sensitivity to inclined magnetic fields with strengths between 1 mG and 50 G, the emergent linear polarization in the 8542~\AA\ and 8662~\AA\ lines is sensitive to magnetic fields in the milligauss range (Manso Sainz \& Trujillo Bueno 2007). 
The reason for this very interesting behavior is that the scattering polarization in the 8498 \AA\ line gets a significant contribution from the selective emission processes that result from the atomic polarization of the short-lived upper level, while that in the 8542~\AA\ and 8662~\AA\ lines is dominated by the selective absorption processes that result from the atomic polarization of the metastable (long-lived) lower levels (Manso Sainz \& Trujillo Bueno 2003b). Therefore, in ``quiet" regions of a stellar atmosphere the magnetic sensitivity of the linear polarization of the 8542~\AA\ and 8662~\AA\ lines is controlled by the lower-level Hanle effect, which implies that in regions with $1\,{\lesssim}\,B\,{\lesssim}\,50$ G the Stokes $Q$ and $U$ profiles are only sensitive to the orientation of the magnetic field vector. In such regions the 
8498~\AA\ line is however sensitive to both the orientation and the strength of the magnetic field through the upper-level Hanle effect. As expected, our multilevel radiative transfer 
calculations for the interpretation of the observations of Fig.~1 show that 
the spatial variations in the observed fractional linear polarization are mainly due to changes in
the orientation of the chromospheric magnetic field.

These types of polarization signal resulting from atomic level
polarization and the Hanle and Zeeman effects can be exploited to
explore the thermal and magnetic structure of the solar
chromosphere. They can also be used to evaluate the degree of realism
of magneto-hydrodynamic simulations of the photosphere-chromosphere
system via careful comparisons of the observed Stokes profiles with
those obtained through forward-modeling.

\subsection{The He~{\sc i} 10830~\AA\ multiplet}

A very suitable diagnostic window for inferring the
magnetic field vector of plasma structures embedded in the solar
chromosphere and corona is that provided by the polarization produced by the joint
action of atomic level polarization and the Hanle and Zeeman effects
in the He~{\sc i} 10830~\AA\ triplet. Off-limb observations give information 
on the magnetic field of spicules and prominences (e.g., Trujillo Bueno et al. 2005; Merenda et al. 2006)  
while on-disk observations show amazing polarization signatures in a variety of plasma structures, such as 
filaments in quiet and active regions (e.g., Trujillo Bueno et al. 2002; Kuckein et al. 2009),
emerging flux regions (e.g., Solanki et al. 2003), sunspots (e.g., Centeno et al. 2006),
flaring regions (e.g., Sasso et al. 2007), etc. The order of magnitude of the ensuing polarization amplitudes varies between 0.1\% and 1\%, approximately, while they are ${\sim}$0.01\% in quiet regions at the solar disk center (see Asensio Ramos et al. 2008). Moreover, Fig.~2 shows a very interesting example of the Stokes profiles observed in polar faculae. The fact that there is a nearby photospheric line of Si {\sc i} whose polarization is caused by the Zeeman effect, makes the 10830~\AA\ spectral region very suitable for investigating the coupling between the photosphere and the corona.
\begin{figure}  
  \centering
  \includegraphics[width=11.5cm]{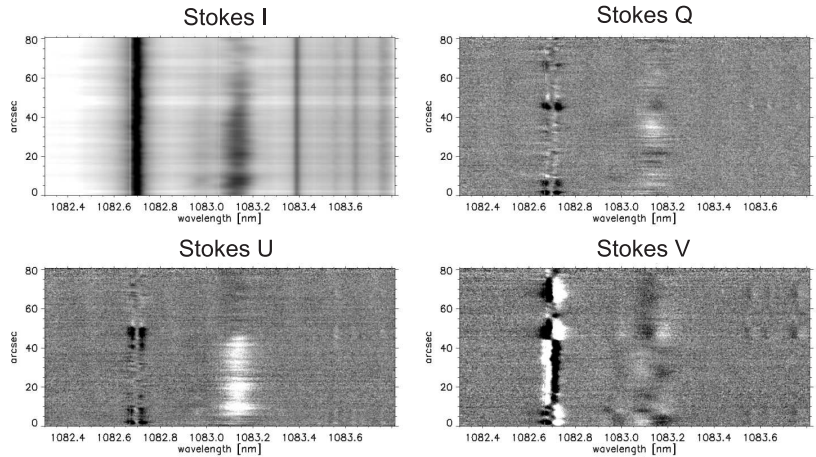}
\caption[]{
  He~{\sc i} 10830~\AA\ spectropolarimetric observation of polar faculae 
  obtained in collaboration with M. Collados (IAC) using the Tenerife Infrared Polarimeter
  attached to the German Vacuum Tower Telescope.
}\end{figure}
For field strengths $B\,{\lesssim}\,100$ G the linear polarization of the He~{\sc i} 10830~\AA\ triplet is fully dominated by the atomic level polarization that is produced by anisotropic radiation pumping processes (Trujillo Bueno et al. 2002). For instance, in Fig.~2 the linear polarization of the He~{\sc i} 10830~\AA\ triplet is mainly caused by atomic level polarization, while the circular polarization is produced by the longitudinal Zeeman effect. 
For field strengths $100\,{<}\,B\,{\lesssim}\,2000$ G the linear polarization of the He~{\sc i} 10830~\AA\ triplet is caused by the joint action of atomic level polarization and the transverse Zeeman effect (Trujillo Bueno \& Asensio Ramos 2007).
This can be seen also in Fig.~2 of Trujillo Bueno (2009), which shows examples of model calculations for the case of a
plasma structure levitating at a height of 2200~km above the
visible solar surface and permeated by a magnetic field of 1200~G with
different orientations. In particular, that figure emphasizes that the influence of atomic level polarization 
on the linear polarization of the He~{\sc i} 10830~\AA\ triplet is very significant, even for magnetic fields as strong as 
1200 G, and that it removes the $180^{\circ}$ azimuth ambiguity present in the Zeeman-effect profiles. 
\begin{table}
\caption{\sffamily {\bf The Ca~{\sc ii} IR triplet}}
\begin{center}
\begin{tabular}{ll}
\hline
Pros & Cons \\
\hline
\\
\parbox[c]{6.2cm}{\small Provides information on the thermal and the magnetic structure, 
all the way up from the photosphere to the bulk of the chromosphere.}  
& \parbox[c]{6.2cm}{\small Not the ideal choice for studying the magnetic field that confines 
the plasma of structures embedded in the solar chromosphere and corona.}\\\\
\parbox[c]{6.2cm}{\small The polarization signals are sensitive to magnetic fields from mG to kG strengths.}  
& \parbox[c]{6.2cm}{\small The forward scattering polarization signals are very weak, 
but nevertheless measurable through longer integration times.}\\\\
\parbox[c]{6.2cm}{\small Good choice to evaluate the reliability of MHD models of the 
photosphere and chromosphere via spectral synthesis and comparison with observations.}  
& \parbox[c]{6.2cm}{\small Stokes inversion of the magnetic field is possible, 
but requires a model for the thermal and density stratification.} 
\\
\hline
\end{tabular}
\end{center}
\end{table}
\begin{table}
\caption{\sffamily {\bf The He~{\sc i} 10830~\AA\ triplet}}
\begin{center}
\begin{tabular}{ll}
\hline
Pros & Cons \\
\hline
\\
\parbox[c]{6.2cm}{\small Good choice for studying the magnetic field that confines the plasma of structures embedded in the solar chromosphere and corona.}  
& \parbox[c]{6.2cm}{\small Not a very suitable choice to study the magnetism of the quiet chromosphere (see, however, \S3.3 and \S5.4 in Asensio Ramos et al. 2008).}\\\\
\parbox[c]{6.2cm}{\small Photospheric lines are 
present in the same spectral region, so information on photospheric magnetic fields  
can also be obtained.}  
& \parbox[c]{6.2cm}{\small It is difficult to obtain information on the thermal and/or density structure.}\\\\
\parbox[c]{6.2cm}{\small Stokes inversion of the magnetic field vector is possible (e.g., via the Hanle+Zeeman code HAZEL developed by Asensio Ramos et al. 2008).}  
& \parbox[c]{6.2cm}{\small Not suitable to evaluate the reliability of MHD models of the solar photosphere and chromosphere.}
\\
\hline
\end{tabular}
\end{center}
\end{table}

\section{Concluding comments}

The exposure time needed for detecting 0.1\% (0.01\%) 
fractional polarization signals with a spectral 
resolution of 50 m\AA\ and a spatial resolution of $1\arcsec$ 
in any of the considered triplets
is of the order of 1~s (60~s), assuming a 1m aperture telescope having  
an overall throughput of 10\%. Tables 1 and 2 summarize the advantages and disadvantages of the information provided by the Stokes profiles produced by atomic level polarization and the Hanle and Zeeman effects in both triplets. While the Ca~{\sc ii} IR triplet is very suitable for exploring the thermal and magnetic structure of the bulk of the solar chromosphere, the He~{\sc i} 10830~\AA\ triplet is the best choice for determining the magnetic field of plasma structures embedded in the solar chromosphere and corona. Obviously, it would be ideal to develop high-sensitivity polarimeters to observe both triplets from a space telescope, but  choosing only one of them would already allow us to discover hitherto unknown aspects of the Sun's hidden magnetism.

\acknowledgements 
Financial support by the Spanish Ministry of Science through project AYA2007-63881 is gratefully acknowledged.


\end{document}